\documentclass[aps,pra,twocolumn,groupedaddress,nofootinbib]{revtex4}

\usepackage{color}
\usepackage[latin1]{inputenc}
\usepackage{amsmath,amscd}
\usepackage{amsfonts}
\usepackage{amssymb}
\usepackage{graphics}
\usepackage{epstopdf}
\usepackage{epsfig}
\usepackage{subfigure}

\usepackage{framed} 

\newcommand{\half}{\frac12}
\newcommand{\eps}{\varepsilon}
\newcommand{\ket}[1]{\left\vert{#1}\right\rangle}
\newcommand{\bra}[1]{\left\langle{#1}\right\vert}
\DeclareMathOperator{\Tr}{Tr}

\newcommand{\kb}[2]{|#1\rangle\langle#2|}
\newcommand{\m}[1]{\mathcal{#1}}

\begin{document}
\bibliographystyle{prsty}
\title{Tomography of a spin qubit in a double quantum dot}

\author{Maki Takahashi}
\author{Stephen D. Bartlett} 
\author{Andrew C. Doherty} 

\affiliation{Centre for Engineered Quantum Systems, School of Physics, The University of Sydney, NSW 2006, Australia}



\date{22 August 2013}

\begin{abstract}
We investigate a range of methods to perform tomography in a solid-state qubit device, for which \textit{\`a priori} initialization and measurement of the qubit is restricted to a single basis of the Bloch sphere.  We explore and compare several methods to acquire precise descriptions of additional states and measurements, quantifying both stochastic and systematic errors, ultimately leading to a tomographically-complete set that can be subsequently used in process tomography.  We focus in detail on the example of a spin qubit formed by the singlet-triplet subspace of two electron spins in a GaAs double quantum dot, although our approach is quite general.
\end{abstract}


\maketitle

\section{Introduction}

Quantum tomography is considered the gold standard for fully characterising quantum systems, and in particular for characterising the quantum logic gates that form the basic elements of a quantum computer.  


In the standard formulation of process tomography~\cite{Chuang97}, a quantum process is characterised through the average statistics of an experiment wherein the unknown process is applied to a system prepared in one of a tomographically complete set of known input states, and the output is subjected to one of a tomographically complete set of known measurements.  Generally, the input states and measurements are assumed to be pure and rank-1, respectively (an approximation that is quite reasonable in a range of optical and atomic systems~\cite{Kwiat,iontrap1}).

This situation in many solid state implementations of qubits is complicated by two issues.  First, one generally does not have access to either a tomographically-complete set of state preparations or measurements in the system; in fact typically, only preparations and measurements in a single basis (say the energy eigenbasis) can be performed directly.  Tomographically complete sets can be generated using transformations (gates) that change these bases, but fully characterising these basis-changing gates through some form of tomographic methods is a bootstrapping problem.  As an illustration of the problem, consider how one would operationally define a direction on the Bloch sphere of a qubit (say, the $x$-basis) in a system where the only preparations and measurements are diagonal in the $z$-basis and where the operations used to rotate into and out of this axis are subjected to both stochastic and systematic errors. 

The second complication is that these state preparations and measurements are often poorly approximated by pure states and projectors.  In many solid state implementations of a qubit, state preparation and measurement (SPAM) errors are significantly larger than gate errors, and so one requires a full characterisation of the SPAM errors prior to performing process tomography.  The effects of SPAM errors, and more broadly the exploration of the effects of systematic and stochastic errors on tomography as well as techniques to overcome these, has been the topic of several recent studies~\cite{Dobrovitski10, Mahler12, Merkel12, Rosset12, Moroder13}.


In this paper, we investigate a number of methods to perform quantum tomography on a system subject to the constraints listed above.  As a specific example we will focus on the singlet-triplet qubit in a double quantum dot~\cite{Petta05,Taylor07,Foletti09}, building on the result of Shulman \textit{et al.}~\cite{Shulman12} by proposing general procedures to perform both state and measurement tomography to characterise these elements for their use in process tomography.  We expect our results to be applicable (possibly with appropriate modifications) to a broader range of solid state systems including other realisations of qubits in semiconductors~\cite{LDqub1,LDqub2,EOqub1,Medford13,Medford13b} and superconductors~\cite{supqub1,supqub2,supqub3}.  Specifically, we introduce three distinct tomographic methods.  The first, Method A, makes additional assumptions about certain state preparations and measurements in order to constrain the problem; these assumptions are unique to our spin qubit, but similar assumptions may be natural in other realisations.  Our second, Method B, uses a phenomenological model for qubit evolution (which incorporates relevant noise processes) to provide further constraints; again, this phenomenological model is specific to spin qubits, but this method could be applied to other qubit realisations for which the evolution is understood phenomenologically.  Finally, our third method, Method C, uses additional state preparations and measurements beyond the `standard' tomographically complete one in order to constrain the problem; we expect this method to be broadly applicable to all qubit systems, and we show that this method has other advantages as well.  Our work complements a number of recent investigations into the tomographic characterisation of spin qubit quantum devices~\cite{Shulman12, Dobrovitski10,Medford13,RohlingBurkard13}.

The outline for this paper is as follows.  In Sec.~\ref{sec:qubit}, we describe the spin singlet-triplet qubit in a double quantum dot that will serve as our model qubit system.  We turn to the construction of tomographically-complete sets of states and measurements in Sec.~\ref{sec:statemeastomo}, where we also provide three distinct recipes for tomography, each with differing sets of assumptions, all of which fully characterise a tomographically-complete set of state and measurement operators.  We compare the convergence of these three recipes. In Sec.~\ref{sec:processtomo}, we numerically study the performance of process tomography using the tomographically-complete sets of states and measurements used in each of the three recipes.  We show, perhaps not surprisingly, that the highest fidelity tomographic reconstructions are obtained using a model for state and measurement tomography with the fewest assumptions.

\subsection{Mathematical elements of quantum tomography}

We briefly review the standard formalism to describe general (mixed) quantum states, generalized quantum measurements, and quantum processes.  See Ref.~\cite{NCQCQI} for details.  For a quantum system with finite-dimensional Hilbert space $\mathcal{H}$, a quantum state is described by a \emph{density matrix} $\rho$, which is a hermitian, positive semi-definite matrix satisfying ${\rm Tr}(\rho)=1$.  Measurements of a qubit are often described as projections along some direction on the Bloch sphere; a description of general noisy measurements includes errors in this measurement direction as well as stochastic errors where the wrong (opposite) direction is identified.  All such measurement errors can be described within the framework of \emph{generalized measurements}, which can describe noisy measurements in the same way that density matrices describe noisy states.  A generalized measurement is formally expressed as a positive operator valued measure (POVM), i.e., a set $\{ E^{(\mu)} \}$ of hermitian, positive semi-definite matrices $E^{(\mu)}$ satisfying $\sum_{\mu=1}^M E^{(\mu)} = I$ where $\mu$ labels the measurement outcome $\mu=1,\ldots, M$ and $I$ is the identity matrix.  Note that, for two outcome measurements, a POVM consists of only two elements, $E^{(1)}$ and $E^{(2)} = I - E^{(1)}$; as such measurements are completely defined by the operator $E^{(1)}$, we can describe the measurement using only this operator, and drop the label $\mu$.  If the state $\rho$ is subjected to a process $\m E$ and then measured with the POVM described by $E$, the probability $p$ of obtaining the measurement outcome associated with $E$ is given by the Born rule $p = {\rm Tr}(\m E(\rho) E)$.  Finally, any quantum process is described by a completely positive (CP) map $\m E$, which maps density operators to density operators, and which should preserve the trace condition.  A unitary evolution $U$, acting on quantum states as $\rho \mapsto U\rho U^{-1}$, is a special case of such a quantum process.

\section{Spin singlet-triplet qubits}\label{sec:qubit}

\begin{figure}
\centering
\includegraphics[width=8cm]{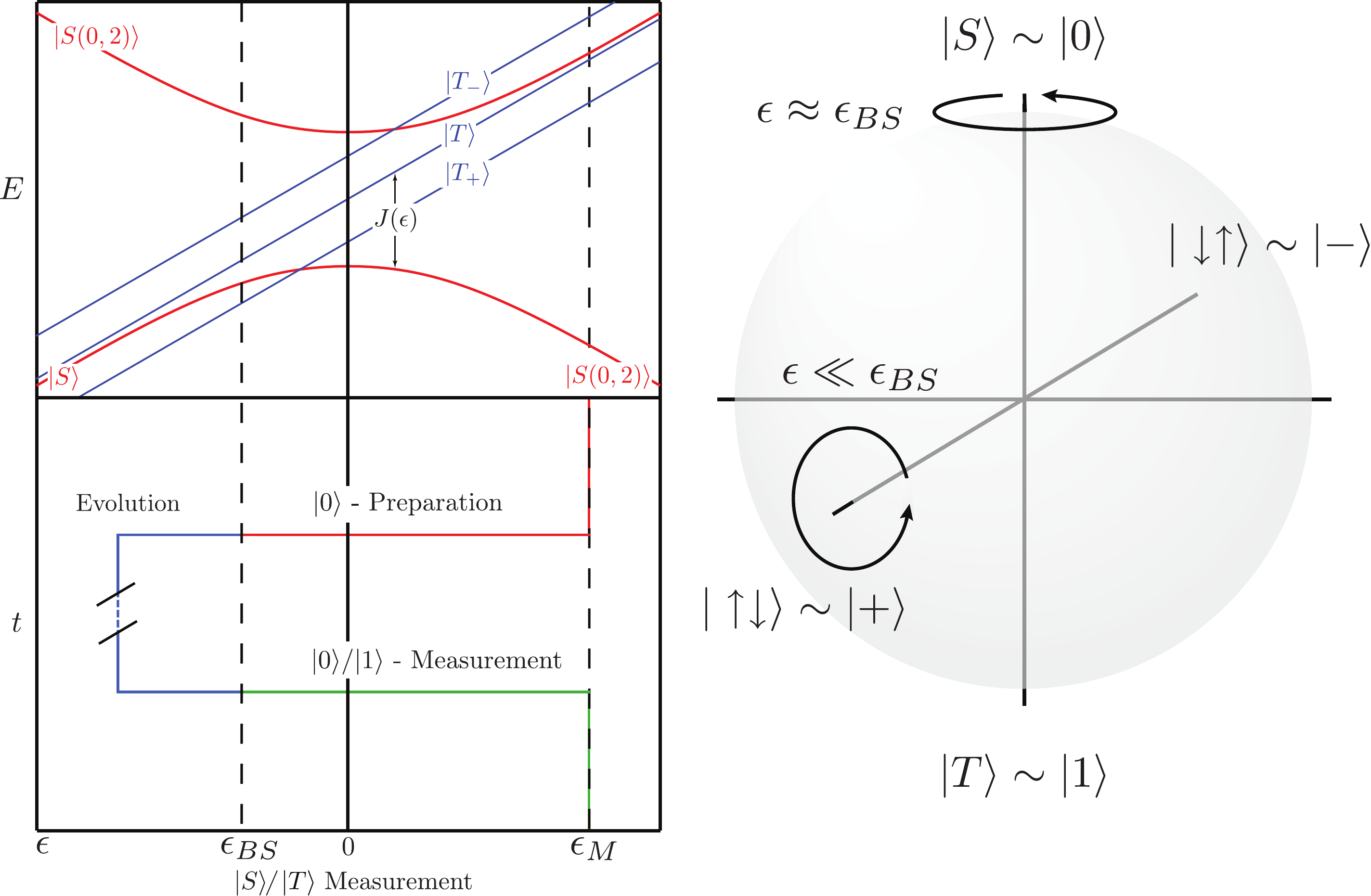}
\caption{The Bloch sphere is defined at the point labelled $\epsilon_{BS}$, where we have plotted the energy levels as a function of $\epsilon$ the gate voltage. The $z$-basis corresponds to the energy eigenstates of the Hamiltonian at the point $\epsilon_{BS}$. \label{bloch}}
\end{figure}

In this section, we briefly review the details of the singlet-triplet spin qubit, following Ref.~\cite{Foletti09}. 
This particular realisation of a qubit consists of the spin states of two electrons trapped in a double quantum dot.  The spin configurations of two electrons each in separate dots spans a four-dimensional space, but a uniform in-plane magnetic field $B$ is applied along the $z$-axis to energetically separate the states $\ket{T_{+}} = \ket{\uparrow\uparrow}$ and $\ket{T_{-}}= \ket{\downarrow\downarrow}$, leaving a two-dimensional space of spin configurations that will define the qubit.  The system is controlled by varying the detuning $\epsilon\propto V_{L}-V_{R}$, where $V_{L}$ and $V_{R}$ are the electrostatic potentials applied to each dot with $L,R$ respectively labelling the left and right dot.  See Fig.~\ref{bloch}.

\subsection{Initialization and readout}

With two electrons in the double dot, initialization of the spin states can be performed using a large bias.   Biasing the potential difference so that $\epsilon>0$ first confines two electrons in one of the two dots with charge configuration $(0,2)$, where the numbers in parentheses labels the occupation number of electrons in the left and right dot respectively. The Pauli-exclusion principle requires that the ground-state wave-function is antisymmetric and hence the spins must be in the singlet state which we label as $\ket{S(0,2)}=\frac{1}{\sqrt{2}}\left(\ket{\uparrow\downarrow} -\ket{\downarrow\uparrow} \right)$. By adiabatically mapping $\epsilon$ from this initialisation point $\epsilon_{M}$ to a point $\epsilon<0$ it is possible to have one electron tunnel to the second dot such that the state $\ket{S(0,2)}\mapsto \ket{S(1,1)} \equiv \ket{S}$ without affecting the spin configuration. 

Readout of the spin state can also be performed by using a large bias together with charge sensing (e.g., via a neighbouring quantum point contact (QPC))~\cite{Barthel09}. By slowly increasing the detuning to a large value $\epsilon_{M}$, the state $\ket{S(1,1)}$ is adiabatically mapped to $\ket{S(0,2)}$, and an electron charge moves from the left to the right dot.  The triplet states do not result in the motion of an electron charge, due to Pauli exclusion.  Thus, by distinguishing these charge configurations with a nearby charge sensor, this spin-dependent charge transfer results in a single-shot measurement of the spin configuration. 


\subsection{The qubit Bloch sphere and Hamiltonian}

The qubit is defined by the two-dimensional space spanned by the $\ket{S}$ and the triplet state $\ket{T_{0}}=\frac{1}{\sqrt{2}}\left(\ket{\uparrow\downarrow} +\ket{\downarrow\uparrow} \right)$.  We note, however, that the charge configuration of these states depends explicitly on the value of the detuning $\epsilon$.  As we change $\epsilon$ we also change the energetics of the system (i.e., we change the Hamiltonian).  It is helpful, and conceptually clearest, to define our qubit and hence our Bloch sphere for a fixed value $\epsilon =\epsilon_{BS}$, where the preparation, evolution and subsequent measurement of our states are all defined relative to this point.  At this point $\epsilon_{BS}$, we will define the $z$-axis of the Bloch sphere to be the energy eigenbasis of the system's Hamiltonian; specifically, $\ket{0} \simeq \ket{S}$ to be the ground state (with singlet character) and $\ket{1} \simeq \ket{T_0}$ the excited state (with triplet character).  These energy eigenstates are not precisely spin singlets/triplets, due to an additional term in the Hamiltonian which we now describe.

Along with the exchange interaction between electrons in the two dots, controlled by the detuning $\epsilon$, the qubit's energetics will be affected by any gradient $\Delta B$ in the magnetic field.  The presence of this gradient field is necessary in order to be able to coherently manipulate the qubit state around two linearly independent directions on the Bloch sphere.  As was demonstrated in Foletti \textit{et al.}~\cite{Foletti09}, the presence of a fixed and constant magnetic gradient field allows arbitrary qubit operations to be performed by simply controlling the detuning parameter $\epsilon$.  
The Hamiltonian for this qubit can be expressed as
\begin{equation}
\hat H(\epsilon) = \tilde J(\epsilon)\hat\sigma_{z} + \tilde{\omega} \hat\sigma_{x}\label{Hamiltonian}
\end{equation}
where $\hat\sigma_{i}$ are the Pauli matrices, $\tilde J(\epsilon) \approx J(\epsilon)+\delta J$ and $\tilde{\omega} \approx \omega+\delta \omega$ for $\omega\equiv g \mu_B \Delta B$. The quantity $\delta J$ represent the errors in the pulses controlling the gate voltages and $\delta \omega$ arises from random fluctuations in the magnetic field gradient.  For each value of $\epsilon$ this Hamiltonian generates rotations approximately around an $\epsilon$-dependent axis at a rate $\Omega \approx \sqrt{J^{2}+\omega^{2}}$. We therefore see that it is possible to achieve arbitrary qubit rotations by simply controlling the gate voltages.  

We see, then, that if the $z$-axis of Bloch sphere is defined by the energy eigenbasis at some value of detuning $\epsilon_{BS}$, the precise spin nature of the energy eigenstates $\ket{0}$ and $\ket{1}$ will be determined by the ratio of $J(\epsilon_{BS})$ and the gradient field $\omega$.  We choose a value of $\epsilon_{BS}$ such that $J(\epsilon_{BS}) \gg \omega$, so that the ground state $\ket{0} \simeq \ket{S}$ has singlet character and $\ket{1} \simeq \ket{T_0}$ has triplet character.  With this choice, the above-described methods for state initialization and readout can accurately prepare and measure in this basis, as states can be quickly mapped to the large value of detuning while still maintaining the adiabatic condition.  We also note that the gradient field implicitly defines the $x$-axis of the Bloch sphere for our qubit; we return to this precise definition later.

\subsection{Types of noise}

It is useful to classify errors associated with our state preparations and measurements into two distinct types:
\begin{itemize}
\item {\it Stochastic errors} resulting from the system coupling to an environment and decohering. 
\item {\it Systematic errors} associated with over- and under-rotations of bases, leading to biases in the state or measurement direction on the Bloch sphere.  (Such errors are sometimes referred to as \emph{unitary errors}.) 
\end{itemize}
For example, in attempting to prepare the $x$-state $\ket{+}$, stochastic errors will result in a mixed state described by a Bloch vector $\vec{r}$ with length $|\vec{r}|<1$, and systematic errors will result in the direction of this Bloch vector being non-parallel to the $x$-axis.  

The systematic errors will be modelled by fixed but randomly determined energies $J$ and $\omega$ in the system Hamiltonian, while stochastic errors will be modelled by white noise fluctuations $\delta J$ and $\delta \omega$ in those same parameters.  Note that, more generally, we could consider a noise spectrum that acts at a range of frequencies, describing a non-Markovian interaction with an environment.  Systematic errors as defined above are the zero-frequency component, whereas Markovian stochastic errors correspond to a white spectrum.  Practically, this would mean that some errors may appear stochastic when describing long-time experiments, but systematic on very short time scales.  A fully general analysis of this situation is beyond the scope of this work, and we restrict our attention to the classification of noise described above.

The stochastic noise associated with $\delta J$ and $\delta \omega$ in the Hamiltonian \eqref{Hamiltonian} results in a decohering map which, due to the white noise approximation, leads to exponential decay in qubit coherence as described by a Bloch equation with a decoherence time $T_2$ determined by the noise power.  The qubit evolution in this case can be represented by a Lindblad master equation which at $\epsilon_{BS}$ takes the form
\begin{equation}
\dot\rho = i\frac{ \Omega} 2\left[ \hat\sigma_{z}, \rho\right] +\frac 1{2 T_{2}}(\hat\sigma_{z}\rho\hat\sigma_{z} - \rho).\label{masterqn}
\end{equation}
The evolution has two parameters, the rotation rate $\Omega$ (which depends on $\epsilon_{BS}$ through the values of $J$ and $\Delta B$ at this point) and the $T_{2}$ decoherence time.

\section{Tomography for states and measurements}\label{sec:statemeastomo}

The standard methods of process tomography make use of a tomographically complete set of states and measurements.  However, as we've described the singlet-triplet qubit so far, we have only discussed how to initialize the qubit in one particular state, and how to measure in a single basis, and in addition both the initialization and readout will be affected significantly by noise.  In this section, we describe how to build up a tomographically complete set of states and measurements using qubit evolutions that introduce both systematic and stochastic errors, and then present tomographic procedures to quantify these errors in a self-consistent way.

\subsection{Tomographically complete sets of states and measurements}

For a single qubit, the standard tomographically-complete set of states and measurements includes preparations of $+x,+y,+z$ eigenstates as well as projective measurements in the $x,y,z$ bases.  As we will make use of noisy (full rank) state and measurement operators, we require an additional state to fix the overall normalization.  We use the additional state $-z$ for this purpose.  In this section, we first discuss noisy preparation and measurement in the $z$-basis, and then further preparations and measurements in different bases to complete a tomographically-complete set of 4 states and 3 measurements.  Using the techniques of Ref.~\cite{Foletti09}, we describe first how to prepare states and measurements that are diagonal in the $z$-basis, then states that are approximately $+x$ eigenstates, and finally describe the most general states and measurements.  

\subsubsection{Initialization and measurement in the $z$-basis}

The initialization process described in Sec.~\ref{sec:qubit} provides our starting point.  As the energy eigenbasis defines the Bloch sphere, there are no systematic errors in this state preparation.  However, we can allow the possibility of stochastic errors due to imprecise relaxation to the ground state.  That is, our initialization leads to a state described as 
\begin{equation}
  \rho_{+z} = (1-\eps)\kb{0}{0}+\eps\kb{1}{1} \label{z_prep}
\end{equation}
where $\eps$ is a free, unknown parameter describing the noise associated with this preparation.  (Recall that $\ket{0},\ket{1}$ are the energy eigenbasis of the qubit at $\epsilon_{BS}$.)

The readout performs a measurement that is also, by definition, along the $z$-axis.  However, it will not in general be well described by a projective measurement, as there will be stochastic errors.  We describe this measurement by the effect (POVM operator) $E_{+z}$, diagonal in the $z$-basis, as
\begin{align}
E_{+z} &= 
    (1- \eps_{0})\kb{0}{0}+\eps_{1}\kb{1}{1} \nonumber\\
                  &= \half(1+(\eps_{1}-\eps_{0}))\hat I + \half (1+(\eps_{1}+\eps_{0}))\hat\sigma_{z} \,. \label{z_meas}
\end{align}
The operator $E_{+z}$ describes the measurement outcome ``$0$'', i.e., the singlet outcome.  (The ``$1$'' measurement outcome is associated with the operator $I - E_{+z}$.)  
The parameter $\eps_{0}$ describes the probability that the measurement will signal the outcome ``$1$'' when the state was actually $\ket{0}$, and $\eps_{1}$ describes the independent probability of signalling the outcome ``$0$'' when the state was actually $\ket{1}$.  

\begin{figure}
\centering
\includegraphics[width=8cm]{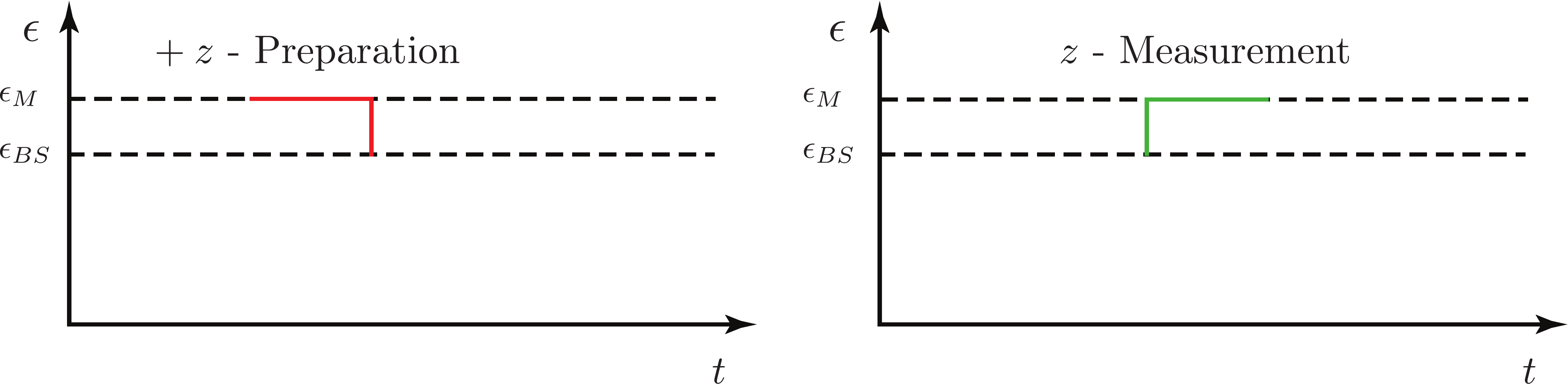}
\caption{Schematic illustration of the $\epsilon$ pulse sequence for the $z$ preparation and measurement as a function of time $t$. The pulse starts at the measurement point $\epsilon_{M}$ and rapidly but adiabatically ramps down to the qubit point $\epsilon_{BS}$.  See Ref.~\cite{Foletti09} for details.\label{z_prep_meas}}
\end{figure}

We therefore have 3 unknown noise parameters, one for $\rho_{+z}$ and two for $E_{+z}$.  However, only two of these are independently observable, even in principle, if $\rho_{+z}$ and $E_{+z}$ are the only states and measurements that can be performed on the double quantum dot system.  (All other states and measurements correspond to a coherent evolution together with these.)  Using this fact and that $\rho_{+z}$ and $E_{+z}$ are both diagonal in the same basis, it is straightforward to show that under an arbitrary evolution of $\rho_{+z}$ it is impossible to distinguish all three parameters.  We can therefore eliminate one of the three noise parameters by redefining the remaining two without affecting the measurement statistics.  Therefore, without loss of generality we will choose $\rho_{+z}$ to be a pure state (that is, choose $\eps = 0$), 
\begin{equation}
\rho_{+z} = \kb{0}{0} = \half\hat I\pm \half\hat\sigma_{z} \,.
\end{equation}
The corresponding measurement will therefore still be of the form \eqref{z_meas} but now with different values for $\eps_0$ and $\eps_1$. We will illustrate the preparation and measurement by the $\epsilon$-pulse sequences illustrated in Fig.~\ref{z_prep_meas}. Note that these parameters 
will in general depend on the choice of $\epsilon_{BS}$.

\subsubsection{Initialization in the $x$-basis}

The magnetic field gradient $\Delta B$ in the Hamiltonian of Eq.~(\ref{Hamiltonian}) provides a different direction on the Bloch sphere than $z$, and allows us to define the $x$-axis of the Bloch sphere as follows.  Performing an initialization where the $|S(0,2)\rangle$ singlet is brought adiabatically to the point where $J(\epsilon)=0$ can be used to prepare an eigenstate of this Hamiltonian term, and rapidly switching the detuning to $\epsilon_{BS}$ completes the initialization; see Fig.~\ref{x_prep_meas}.  We define this state $\rho_{+x}$ to be real, i.e., its Bloch vector lies in the $x{-}z$ plane; in general, systematic errors due to the implementation of the control pulse will mean this state is not precisely aligned with the $x$-axis.  We can therefore represent $\rho_{+x}$ by
\begin{align}
  \rho_{+x} &= \half I +\half \left(r_{x}^{(2)} \sigma_{x} + r_{z}^{(2)}\sigma_{z}\right). \label{x_prep}
\end{align}
where $r_{x}^{(2)}$ and $r_{z}^{(2)}$ are the Bloch sphere components constrained such that $(r_{x}^{(2)})^{2}+(r_{z}^{(2)})^{2}\leq 1$.  (Here, the superscript $(2)$ denotes that this is our second independent preparation.)  We emphasise that the above equation is a completely general expression for the form of the $+x$ preparation, regardless of the specific method (pulse sequence) used to generate it.


\subsubsection{The remaining states and measurements}


For process tomography, we require at least four states $\rho_i$, $i=1,2,3,4$ and three measurements $E_j$, $j=1,2,3$.  With $\rho_1 = \rho_{+z}$, $\rho_{2} = \rho_{+x}$ and $E_{1} = E_{+z}$ as defined above, we still require at least two or more additional states $\rho_{i}$ for $i=3,4,\ldots$, and at least two additional measurements $E_{j}$ for $j=2,3,\ldots$.  In general, these states will have both stochastic and systematic errors.  

Additional states can be initialized by reducing $J(\epsilon)$ to zero, or any nonzero value, and allowing the qubit to evolve prior to switching back to the point $\epsilon_{BS}$.  Rather than attempting to describe the effect of such pulse sequences, we leave the form of these initializations completely general, represented as an arbitrary qubit density operator as
\begin{equation}
\rho_{i} = \half I  + \frac 12 \sum_{a= 1}^{3} r_{a}^{(i)}\sigma_{a}\label{arb_state}
\end{equation}
where each state $i$ has three unknown parameters $r^{(i)}_{a}$ for $a = 1,2,3$ corresponding to the $x$, $y$, and $z$ components of the Bloch vector. The value in parentheses labels the different state preparations, $i=3,4,\ldots$.  

\begin{figure}
         \subfigure[]{
	 \centering
	  \includegraphics[width=8cm]{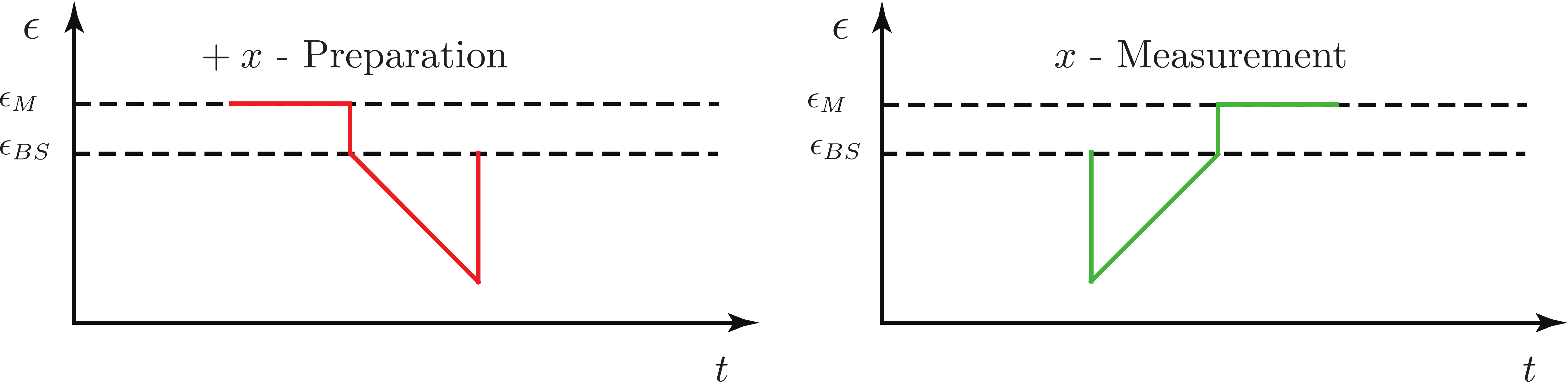}
	  \label{x_prep_meas}
	}
	\subfigure[]{
	 \centering
	  \includegraphics[width=8cm]{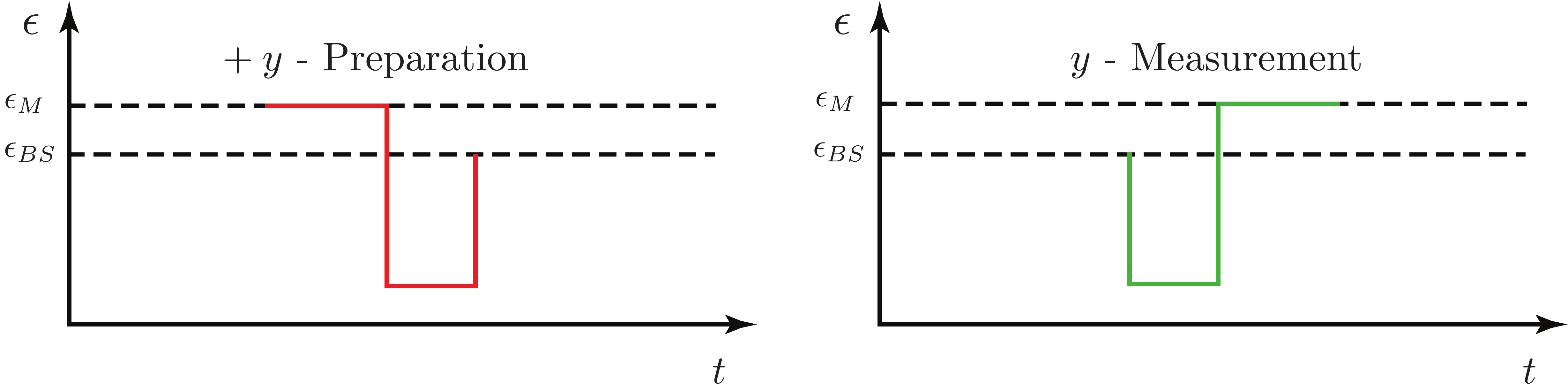}
	  \label{y_prep_meas}
	}
\caption{Schematic illustration of the pulse sequence for the \subref{x_prep_meas} $+x$ and \subref{y_prep_meas} $+y$ preparation and measurement.\label{prep_meas}}
\end{figure}

Similarly any additional measurements are also left completely general and are explicitly represented as
\begin{equation}
  E_{j} = \half \left( 1- (\eps_{1}-\eps_{0}) \right) I +\half (1-(\eps_{1}+\eps_{0})) \sum_{a=1}^{3}R_{a}^{(j)}\sigma_{a}\label{arb_meas}
\end{equation}
where the unknowns are the 2 noise parameters $\eps_{0,1}$, the three measurement parameters $R^{(j)}_{a}$ for $a=1,2,3$, and again where the value in parentheses labels the different measurements $j=2,3,\ldots$.  Note that we retain the $z$-axis measurement noise parameters $\eps_{0,1}$, because any measurement on this system correspond to an evolution subsequently followed by the original $E_{+z}$ measurement.  However, additional stochastic noise is included in this description as well, represented by the possibility that $\sum_{a=1^3} R^{(j)}_a$ can be less than unity.  In order for these to correspond to physical states and  measurements they must be constrained such that $\sum_{a=1}^{3}r^{(i)}_{a} \leq 1$ and $\sum_{a=1}^{3}R^{(j)}_{a} \leq 1$, where equality denotes pure states and projective measurements. The pulse sequences for these states and measurements are illustrated in Fig.~\ref{prep_meas}. 

\subsection{Characterising the states and measurements}

The task of characterising the states and measurements is now reduced to simply determining the unknown parameters for an informationally complete set of states and measurements.  In order to identify these unknowns, one would collect statistics from preparing and then measuring in all combinations of states and measurements, and then fit the unknown parameters to the data for example by using maximum likelihood estimation (MLE).  Unfortunately, to characterise a minimal set, i.e., only 4 states and 3 measurements, we have 16 unknown parameters to determine, and the statistics from measuring these 4 states with 3 measurements will only give 12 independent pieces of data.  We therefore will either require additional assumptions to restrict the number of free parameters, or additional states and measurements to provide more independent statistics.  

We define and compare three different methods for this, each based on a different approach. 

\subsubsection{Method $A$ -- additional assumptions}

We reduce the number of unknown parameters by making additional assumptions about the form of the states and measurements.  One natural assumption is that the measurements $E_{2,3}$ do not have any additional stochastic errors apart from those due to the underlying $E_{+z}$ measurement, described by $\eps_{0,1}$.  (The measurements can still possess arbitrary systematic errors.)  This assumption corresponds to enforcing the additional constraints $\sum_{a=1}^{3}(R^{(j)}_{a})^{2} =1$ for $j=2,3$ which eliminates 1 unknown for each measurement, yielding a total of 13 unknown parameters in the model.  Another natural assumption is that the measurement $E_2$, which uses an inverted pulse sequence to that used in the $\rho_{+x}$ initialization, is similarly constrained to the $x-z$ plane on the Bloch sphere, i.e., enforcing $R^{(2)}_3 = 0$.  This additional assumption reduces the total number of unknown parameters in the model to 12, precisely equal to the number of independent data we can obtain.

We note these assumptions are particular to our singlet-triplet qubit system.  To apply this method to other qubit realisations, one would need to identify a similar number of `natural' assumptions to constrain the problem, and we note that this may not be possible in all situations.

\subsubsection{Method $B$ -- using free evolution}

As an alternative method, we can avoid making additional assumptions on the form of our states and measurements, and follow the general approach of Shulman \textit{et al.,}~\cite{Shulman12}.  In this approach, we obtain additional data by allowing the qubit to evolve freely for some time between the state initialization and the measurement.  By fitting the evolution to a simple phenomenological model with few free parameters, we can constrain the parameters of all states and measurements in our tomographically-complete set.

We consider allowing the qubit to evolve freely for various times between the preparations and measurements.  This will introduce additional unknown parameters associated with the evolution.  In order to reduce the number of unknown parameters introduced, we will evolve the states under the Hamiltonian defined at the qubit point, i.e., at $\epsilon_{BS}$, and to a good approximation the evolution corresponds to rotations around the $z$-axis with some decay towards the $z$-axis. Specifically, we model evolution of an arbitrary quantum state $\rho$ by the Lindblad master equation of Eq.~(\ref{masterqn}).  This model introduces two unknown parameters: the rotation rate $\Omega$ and the $T_{2}$ coherence time. The solution is given by
\begin{multline}
\rho(t)  = \tfrac{1}{2} \left(1+ e^{-t/ T_{2}}\cos \Omega t \right)\rho(0) \\
+ \tfrac{1}{2} \left(1- e^{-t/ T_{2}}\cos  \Omega t \right)\sigma_{z}\rho(0)\sigma_{z} \\
+\tfrac{i}{2} e^{-t/ T_{2}}\sin \Omega t  \left[ \sigma_{z}, \rho(0)\right].
\end{multline}
We will take the above form of $\rho(t)$ as an ansatz for the state obtained by initializing the qubit as $\rho(0)$ and then allowing it to evolve freely for time $t$.  
The conditional probability $ p_{j|i} = \m P(E_{j}|\rho_{i}(t))$ for obtaining the measurement outcome $E_{j}$ given the initial state $\rho_{i}$ and an evolution time $t$ takes the form
\begin{align}
  p_{j|i}(t) &=  a_{i,j} +e^{-t/ T_{2}}\left(b_{i,j} \cos  (\Omega t)- c_{i,j} \sin  (\Omega t)\right)   \label{con_prob}
\end{align}
We note that this method can be modified, in particular in the form of the decoherence, if the qubit evolution is more accurately modelled by an alternative parametrisation.  For example, in a non-Markovian noise environment, other non-exponential decay envelopes may be more appropriate.  We emphasise that any such model forms an assumption on which this method is based.

With this evolution, we can constrain all of the parameters in our tomographically-complete sets of states and measurements, as well as the additional free parameters in the above evolution, by allowing the qubit to evolve for a number of finite timesteps $t_k$ between the various qubit initializations and the subsequent measurements.  We note, however, that this method is based on the assumption of qubit evolution according to the above ansatz.  This Method B can be adapted to other qubit systems using a phenomenological evolution appropriate to that qubit and its noise environment.

\subsubsection{Method $C$ -- more states and measurements}

Finally, as a third method, we investigate how we can add additional states $\rho_{i}$ and measurements $E_{j}$ beyond the minimal tomographically complete set in order to further constrain the parameters of our fit.  The number of unknown parameters introduced by adding $K$ additional states and measurements grows linearly with $K$; specifically, we introduce $6$ new parameters for each state and measurement pair using the completely general forms of \eqref{arb_state} and \eqref{arb_meas}.  Performing all possible combinations of state initialization and measurement in an experiment, the amount of independent data grows quadratically in $K$.  Therefore, provided we introduce a sufficient numbers of state and measurement pairs we can always collect a sufficient set of independent data to determine all unknown parameters without the need of evolving our states or introducing any other assumptions.  In our case, it suffices to make use of a total of 5 state and measurement pairs:  that is, $4$ states and $3$ measurements from our tomographically-complete sets, plus $1$ more arbitrary state and $2$ more arbitrary measurements.  Therefore, we would have $25$ unknown parameters and exactly $25$ independent measurements with the probabilities given by $ p_{j|i} = \m P(E_{j}|\rho_{i})$.

We emphasise that, unlike methods $A$ and $B$ above, this method introduces no additional assumptions; beyond the  $z$-basis state and measurement, all other states and measurements are completely free and arbitrary.  As a result, Method C is broadly applicable to all qubit realisations that possess a common $z$-basis state preparation and measurement. 



\subsection{Tomography and the state and measurement reconstructions}

We now describe how experimental data following the above three methods can be used to reconstruct the parameters of the states and measurements, therefore providing an accurate estimate of these states and measurements for future use in process tomography.  

Let us first consider Method $B$, which requires qubit evolution.  We select $M$ equally spaced intervals of time $t_k$, $k=1,\ldots,M$.  For each state $i=1,2,3,4$, measurement $j=1,2,3$, and time $t_k$, $k=1,\ldots, M$, we generate a set of data corresponding to the conditional probabilities \eqref{con_prob} which corresponds to preparing the state $\rho_{i}$ evolving for a fixed period of time $t_k$ and measuring $E_{j}$.  This is repeated $N$ times with $n_{j|i}(t_k)$ positive outcomes.  The statistics are labelled by the measurement $i$, the initial state $j$ and the time $t$ with a frequency $\tilde p_{j|i}(t_k) = n_{j|i}(t)/N$.  The 12 data sets are then collected and likelihood function $\m L$ for the 12 conditional probabilities \eqref{con_prob} is maximized over the parameter space to determine the best fit. Although $\tilde p_{j|i}(t_k)$ will be distributed binomially, for sufficiently large $N$ we can approximate this by a normal distribution.  The log-likelihood becomes  
\[
-\ln\m L = \sum_{i,j}\sum_{k=1}^{M}\frac1{\sigma_{ijk}}\left({\rm Tr}(E_{j}\rho_{i}(t_{k})) -\tilde p_{j|i}(t_{k})\right)^{2}
\]
where $\sigma_{ijk} = \sqrt{N \tilde p_{j|i}(t_k) (1- \tilde p_{j|i}(t_k))}$ are the errors associated with the measurements. Minimizing $-\ln\m L$ we obtain the best fit values of the 16 unknown parameters characterising the states and measurements and the two evolution parameters $\Omega$ and $T_{2}$. Note that if the reconstructed states or measurements were found to be unphysical then they should be corrected for in the standard manner \cite{Kwiat}.

For methods $A$ and $C$ the situation is considerably simpler.  For both methods, as the measurement frequencies $\tilde p_{j|i} = n_{j|i}/N$ completely constrain the unknown parameters, we similarly construct $-\ln\m L$ and minimize over all parameters to determine the states. 

We emphasize that the fit in all cases is of the conditional probabilities $p_{j|i} = {\rm Tr}(E_j \rho_i)$, which are non-linear functions of the unknown parameters.  Therefore, to achieve a good fit, in practice we require a reasonable initial estimate.


\subsection{Simulations of state and measurement tomography}

We now will explore how the above methods behave using simulated data, and compare the relative accuracy and convergence of the three methods.

The assumptions used in these methods provide a fundamental problem in doing simulations.  Both Methods $A$ and $B$ make explicit use of assumptions:  Method $A$ makes assumptions about some of the state initializations and measurements, and Method $B$ makes assumptions about the free evolution of the qubit.  In our simulations (as with actual experiments) the performance of these methods will obviously depend on the accuracy of these assumptions.  In the following, we describe simulations for which the assumptions of Method $A$ are explicitly violated, but the assumptions of Method $B$ are obeyed, and we compare these methods in light of this.

In our simulations, the values of the unknown parameters in the states and measurements were selected with systematic errors described by inaccuracies in the measurement directions on the Bloch sphere of $\sim 10^{\circ}$ and stochastic errors $\sim0.05$, which roughly corresponds to those found in Ref.~\cite{Foletti09}.  Note the results presented here in do not depend significantly on the particular choice of parameters.  In particular all the measurements have additional noise, and the $x$-measurement $E_{2}$ is not constrained to lie in the $x{-}z$ plane as assumed by Method $A$.  Also in our simulations, qubit evolution is modelled by the Lindblad master equation of Eq.~(\ref{masterqn}); therefore this assumption in Method $B$ is precisely obeyed in our simulation.  The minimization of the log likelihood is performed using \emph{Minuit}, a minimization routine in the ROOT library developed by CERN.

\begin{figure}
        \subfigure[]{
	 \centering
	  \includegraphics[width=8cm]{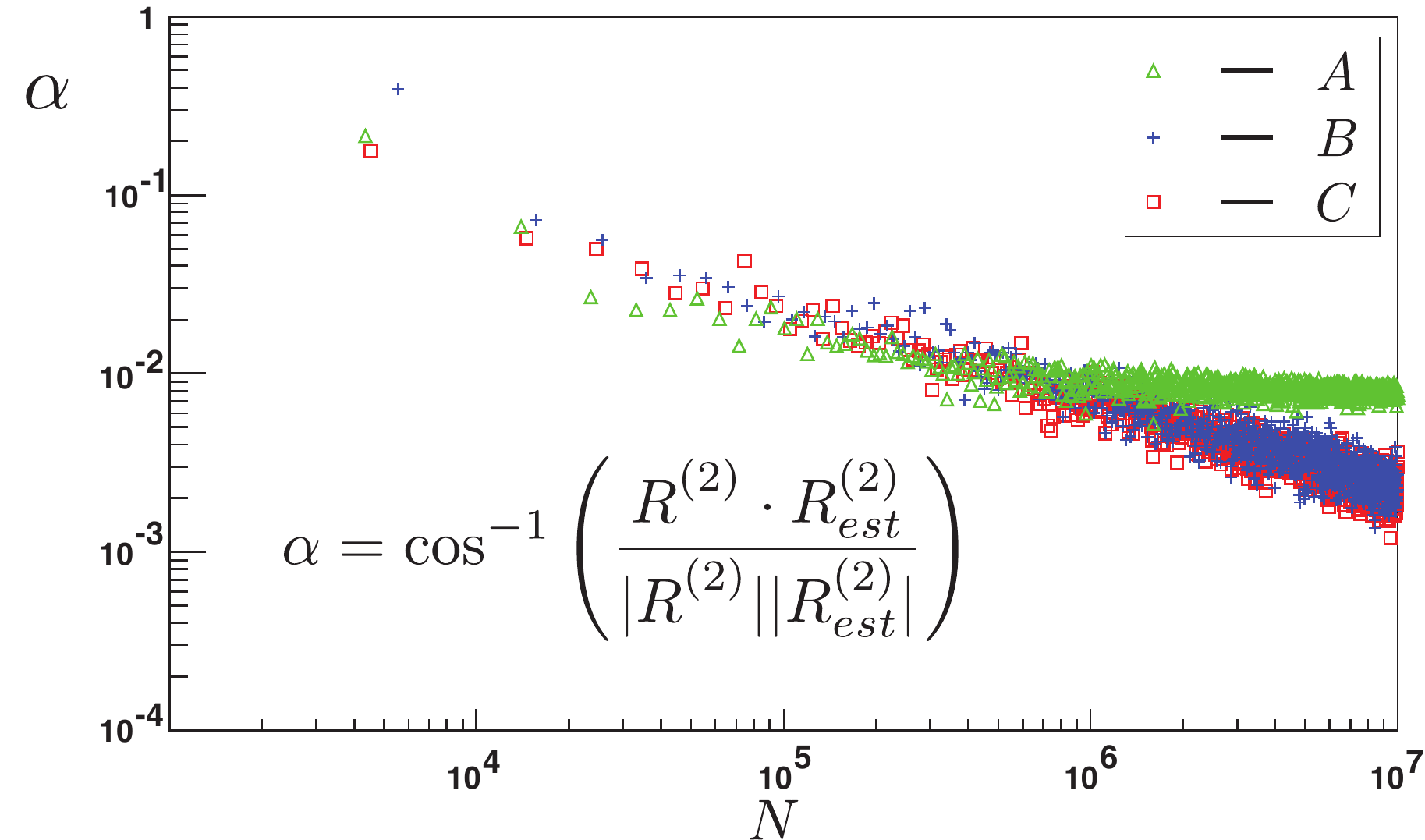}
	  \label{ang_distance}
	}
	\subfigure[]{
	\centering
	  \includegraphics[width=8cm]{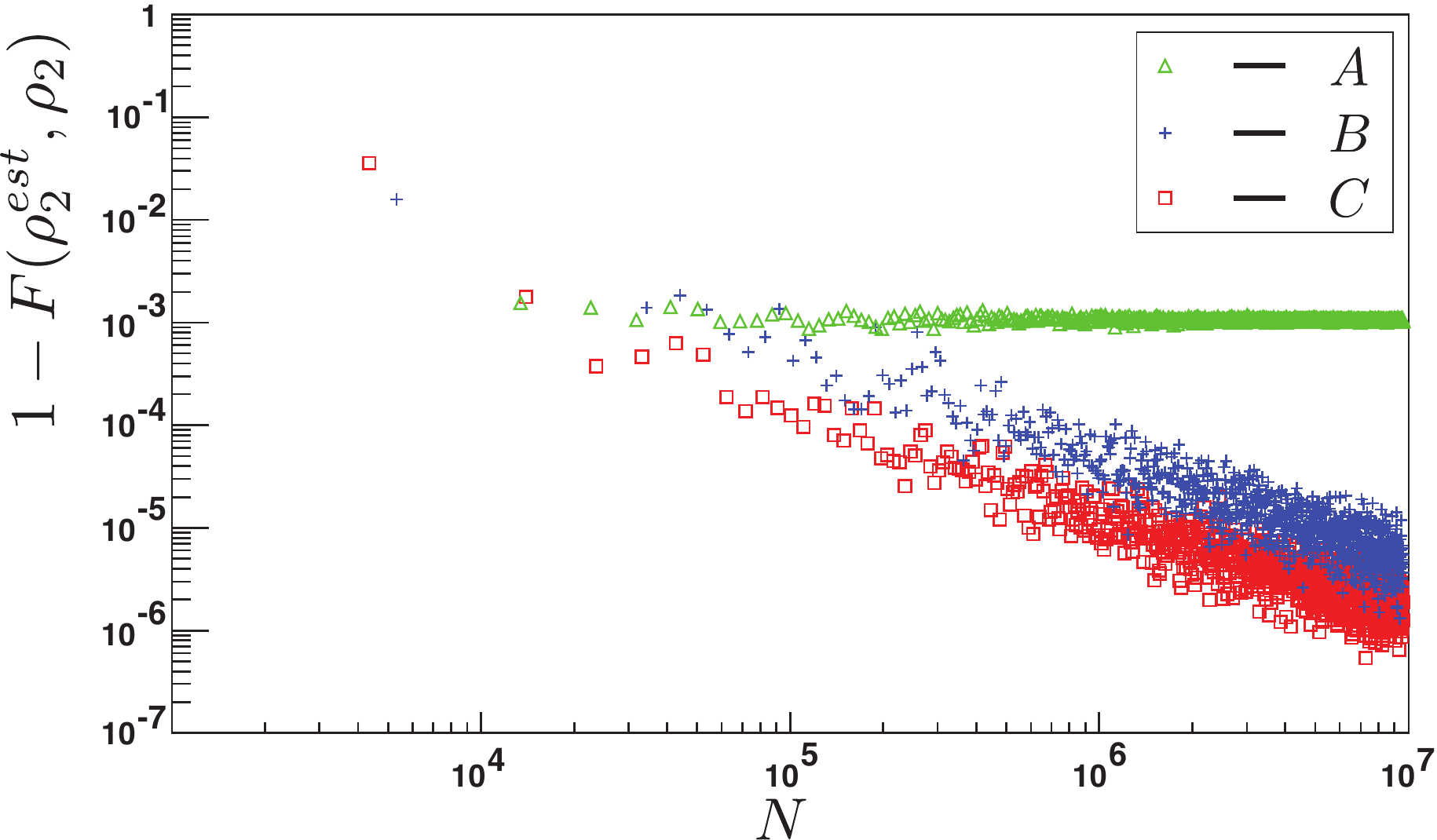}
	  \label{state_fidelity}
	}
	\caption{Comparison of state and measurement reconstruction for method $A$ (green), $B$ (blue) and $C$ (red). \subref{ang_distance} The convergence of a systematic error for the measurement $E_{x}$. Here, $\alpha$ is the angular separation between the vector $R^{2}_{i}$ and the true measurement with respect to the total number of measurements $N$. \subref{state_fidelity} The infidelity $1-F$ between the true state $\rho_{1}$ and reconstructed state $\rho^{est}_{1}$.  The data has been averaged over 10 runs per point in order to highlight the difference in average-case performance of Methods $B$ and $C$.} \label{comp_data}
\end{figure}

If we compare how quickly the systematic errors for the measurements converge to the true values, Fig.~\ref{ang_distance}, we find that at least for sufficiently small stochastic errors in the true measurements ($\lesssim0.05$) the systematic errors converge to the true errors at exactly the same rates for $N\lesssim10^{6}$.  However, for $N$ larger we see that method $A$ obtains a lower limit on the accuracy of the systematic errors.  We find through simulations that this lower limit on the accuracy is primarily due to the inaccurate assumption about the stochastic errors on measurement $j=2,3$ and not the inaccurate assumption that the $x$-measurement lies in the $x{-}z$ plane.

In Fig.~\ref{state_fidelity}, we compare the convergence of the reconstructed states to the true states, as quantified by the fidelity $F(\rho_{i}^{est},\rho_{i})$.  For both methods $B$ and $C$ we find that the reconstructed state converges to the true state at a rate $\propto N^{-1}$. For method $A$ we find that for small $N\lesssim100$ the reconstructed state converges at approximately the same rate, however for $N$ larger we find that the fidelity reaches a lower bound of the order $10^{-3}$. This lower limit is set by the stochastic errors associated with the measurements which were assumed to be sufficiently small that they can be ignored.  We note that Method $C$ performs better \emph{on average} than $B$ by a constant factor (about a factor of $5$), although the spread in this performance is significant.  (Note that the data in Fig.~\ref{state_fidelity} has been averaged over 10 runs per point to highlight this difference in average-case performance.)
 
As mentioned above, given that the conditional probabilities $p_{j|i} = {\rm Tr}(E_j \rho_i)$ are non-linear functions of the unknown parameters, to achieve a good fit we in practice require a reasonable initial estimate.  Here, we found that method $B$ was the most stable and in practice one could be completely ignorant about all of the parameters and still achieve a reasonable fit to the data. This is primarily because the extra evolutionary degrees of freedom reduce the parameter space compatible with the data.  (This stability suggests that one may wish to increase the number of states and measurements in method $C$ beyond the minimum required, in order to obtain similar stability.)  For methods $A$ and $C$ this is not the case, and we require a reasonable approximation to the true states and measurements, for example we found that it was necessary to take $\rho_{2}$ to be approximately $\kb{+}{+}$, likewise for any additional states in order to achieve a reasonable fit. Finally method $A$ was the most sensitive to the initial conditions and for a fair comparison we have plotted the data with initial parameters to be within $1\%{-}2\%$ of their true value.
    
\section{Process tomography}\label{sec:processtomo}

The methods described in the previous section provide estimates (reconstructions) of a tomographically-complete set of states and measurements, which can subsequently be used to perform process tomography.  Here, we investigate the performance of process tomography based on these reconstructions.  

As an example, we consider process tomography of a noisy implementation of the Hadamard gate.  The Hadamard gate is unitary, but any experimental implementation will have both stochastic and systematic errors. The resulting evolution is therefore a CP map $\m E$, which we would like to estimate.  We will use our tomographically-complete sets of states $\{\rho_i\}$ and measurements $\{ E_j \}$, reconstructed to some accuracy as parameterised by the number of measurements $N_{\rm SPAM}$.  Process tomography then follows the standard procedure \cite{NCQCQI}, however with the added complication that the states and measurements are nonorthogonal and noisy.  

For each state $i$ and measurement $j$, we collect statistics from $N$ experiments where the input state $\rho_i$ is acted upon by the gate $\m E$ and subsequently measured with $E_j$.  The relative frequencies, $\tilde p_{j|i} = n_{ij}/N$, which are estimates of the quantum mechanical probabilities $p_{j|i} = p(E_{j}|\m E(\rho_{i}))$.  Given our estimates for $\rho_i$ and $E_j$ obtained through state and measurement tomography, we can reconstruct the process $\m E$.

For the numerical implementation of this reconstruction, it is useful to consider the Choi state $\rho_{\m E_{\rm rec}}$ associated with this process.  Recall that, for a qubit, the Choi state for a process $\m E$ is given by $\rho_{\m E} = [\m I\otimes\m E](|\Phi^+\rangle\langle\Phi^+|)$, where $|\Phi^+\rangle = 1/\sqrt{2}(\ket{00}+\ket{11})$.
In practice this simple inversion using the Born rule will typically yield an unphysical Choi state.  A standard method to obtain only physical Choi states is to perform MLE on the reconstructed state $\rho_{\m E_{rec}}$ constrained to the set of physical Choi states. There are two conditions that a physical Choi state must satisfy
\begin{enumerate}
 \item it must be an Hermitian, $\Tr\rho_{\m E}=1$, positive semi-definite operator, i.e., $\bra{\psi}\rho_{\m E}\ket{\psi}\ge 0$ for any $\ket{\psi}$. 
 \item The partial trace over subsystem $B$ should yield the maximally mixed state, i.e., $\Tr_{B}(\rho_{\m E}) = \half I_{A}$.
 \end{enumerate} 
The first condition will be enforced by choosing a suitable parametrization for positive semi-definite states. Namely, the Cholesky decomposition where $\rho_{\m E} = T^{\dagger}T/ \Tr(T^{\dagger}T)$ where $T$ is a lower triangular complex matrix which has $16$ degrees of freedom $t_{i}$, $i = 1,2,\ldots,16$ \cite{Kwiat}. The second condition can now be expressed as a set of 4 constraints on the parameters $t_{i}$, $C_{i}(t_{i})=0$, for $i=1,\ldots,4$.   

In order to perform maximum likelihood estimation, we construct the likelihood function $\m L$ by considering an operator basis of $16$ elements formed by the tensor product $M_{ij} = M_{i}\otimes M_{j}$ where $M_{i}$ are an operator basis for a qubit (for example, the Pauli spin matrices).  We define $p_{ij}$ and $q_{ij}$ to be tho components of the reconstructed operator $\rho_{\m E_{\rm rec}}$ and the physical Choi state $\rho_{\m E_{\rm est}}$. The resulting log-likelihood function is
\[
-\ln\m L =N\sum_{ij}\frac{(p_{ij}-q_{ij})^{2}}{q_{ij}(1-q_{ij})}
\]     
where $p_{ij} = \Tr(M_{ij} \rho_{\m E_{\rm rec}})$ and $q_{ij} = \Tr(M_{ij} \rho_{\m E_{\rm est}})$. As is standard practice we have ignored the normalisation constant which enters into the likelihood function~\cite{Kwiat}. We would now like to minimize $-\ln\m L$ over the parameters $t_{i}$ whilst enforcing the constraints $C_{i}=0$. In order to enforce these constraints we will follow a procedure similar to that of Ref.~\cite{obrien}. We can turn this into an unconstrained problem by following what is called the {\it augmented Lagrangian method} where we add a penalty function to $-\ln\m L$ \cite{Optimization} 
\begin{equation}
L(t_{1},t_{2},\ldots) = -\ln\m L+\sum_{i=1}^{4} \left(\lambda_{i} C_{i}+ \frac \mu 2 C_{i}^{2}\right)
\end{equation}
where in optimization theory $L$ is called the Lagrangian function. The minimization of $L$ follows an iterative procedure where $\lambda_{i}$ and $\mu$ are appropriately chosen constants which are updated after each minimization. The advantage of this method is that the solutions do not depend strongly on the initial guess and will converge to the minimum in a finite number of iterations (here we found that 5 iterations were sufficient).  Minimizing $L$ over the set of parameters $t_{i}$ results in the closest physical Choi state consistent with the data satisfying conditions 1 and 2~\cite{trace_condition}.

\begin{figure*}
        \subfigure[]{
	 \centering
	  \includegraphics[width=8cm]{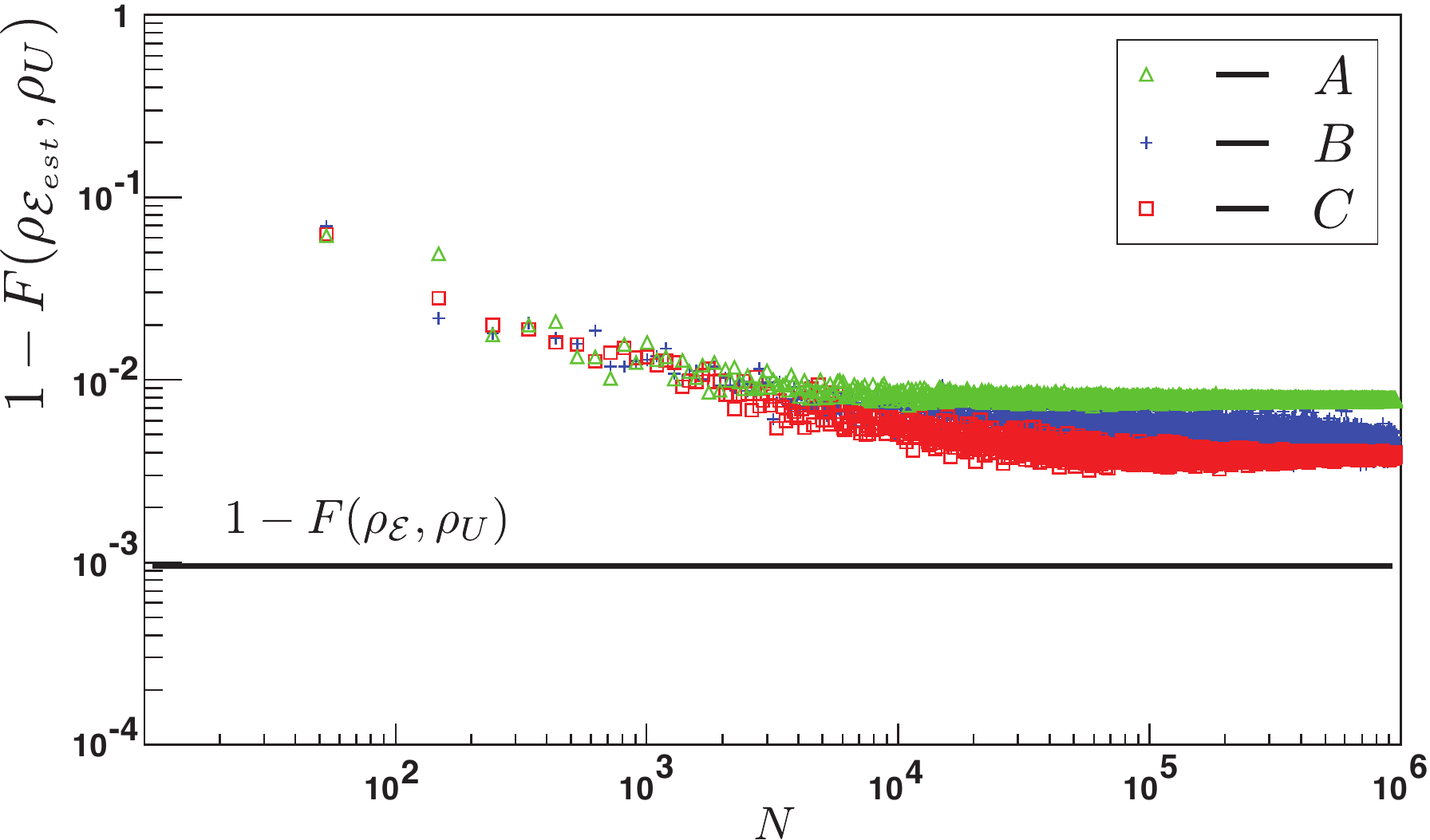}
	  \label{process_fidelity_ideal_sm1000}
	}
        \subfigure[]{
	 \centering
	  \includegraphics[width=8cm]{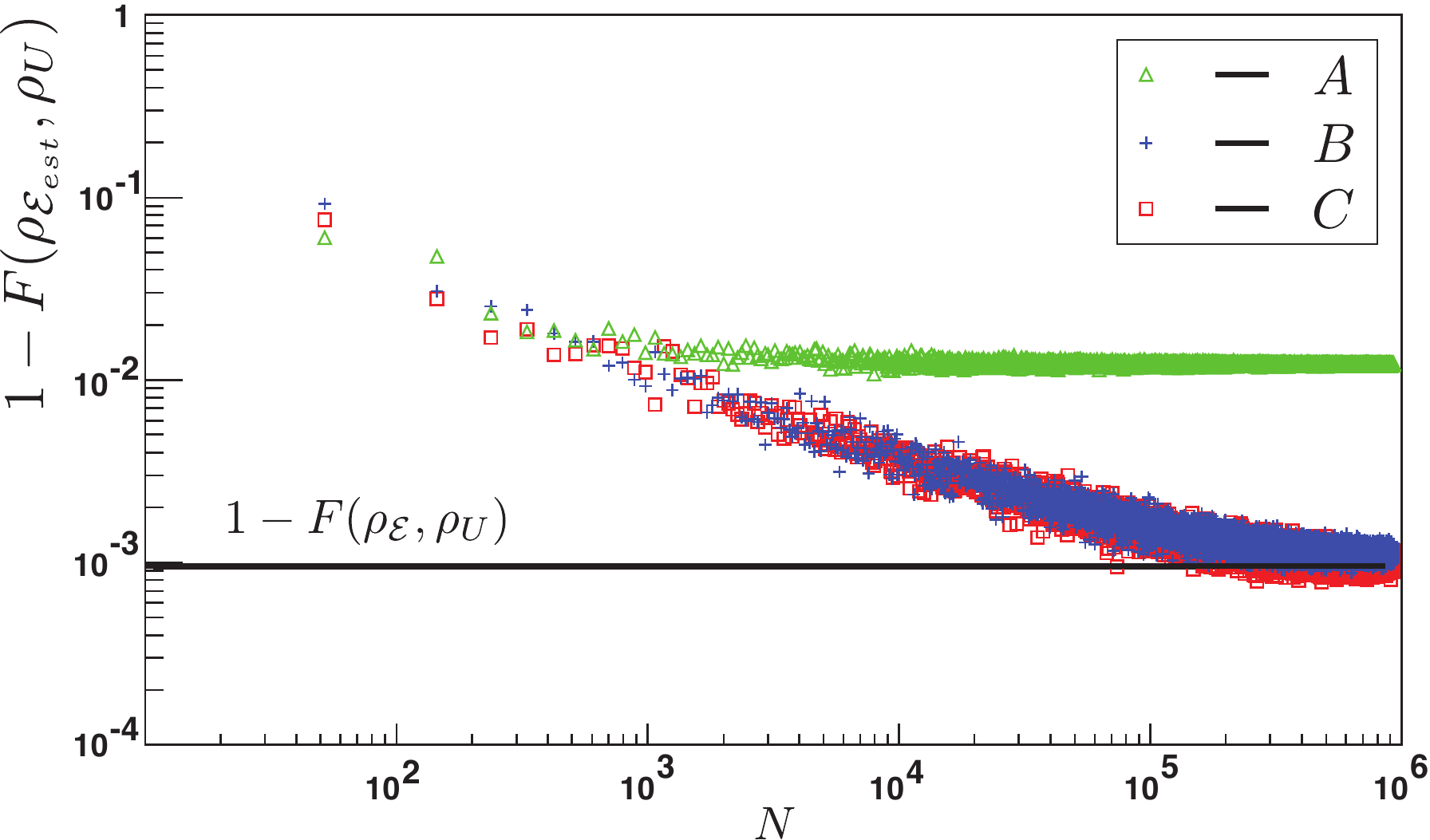}
	  \label{process_fidelity_ideal}
	}
	\subfigure[]{
	\centering
	  \includegraphics[width=8cm]{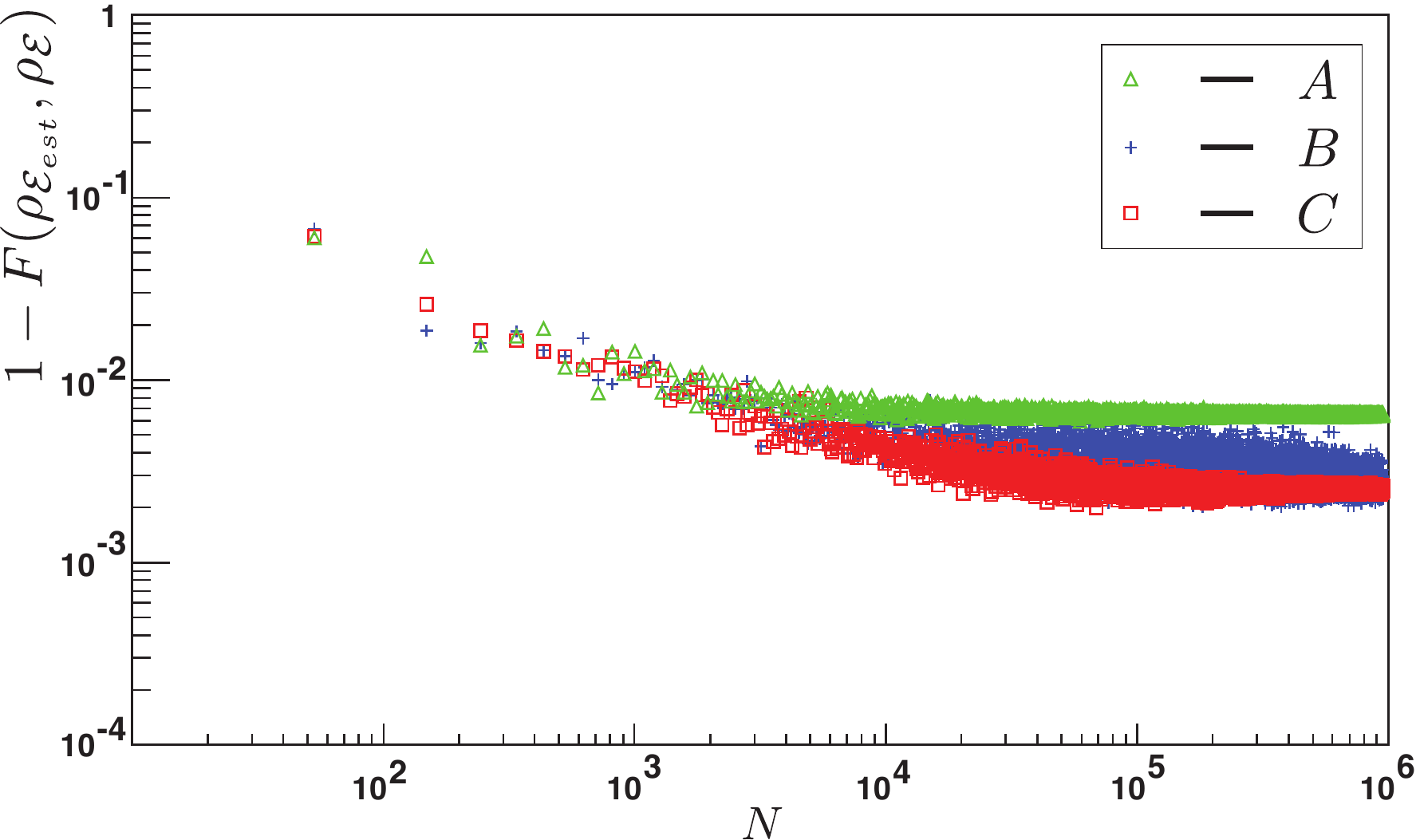}
	  \label{process_fidelity_true_sm1000}
	}
	\subfigure[]{
	\centering
	  \includegraphics[width=8cm]{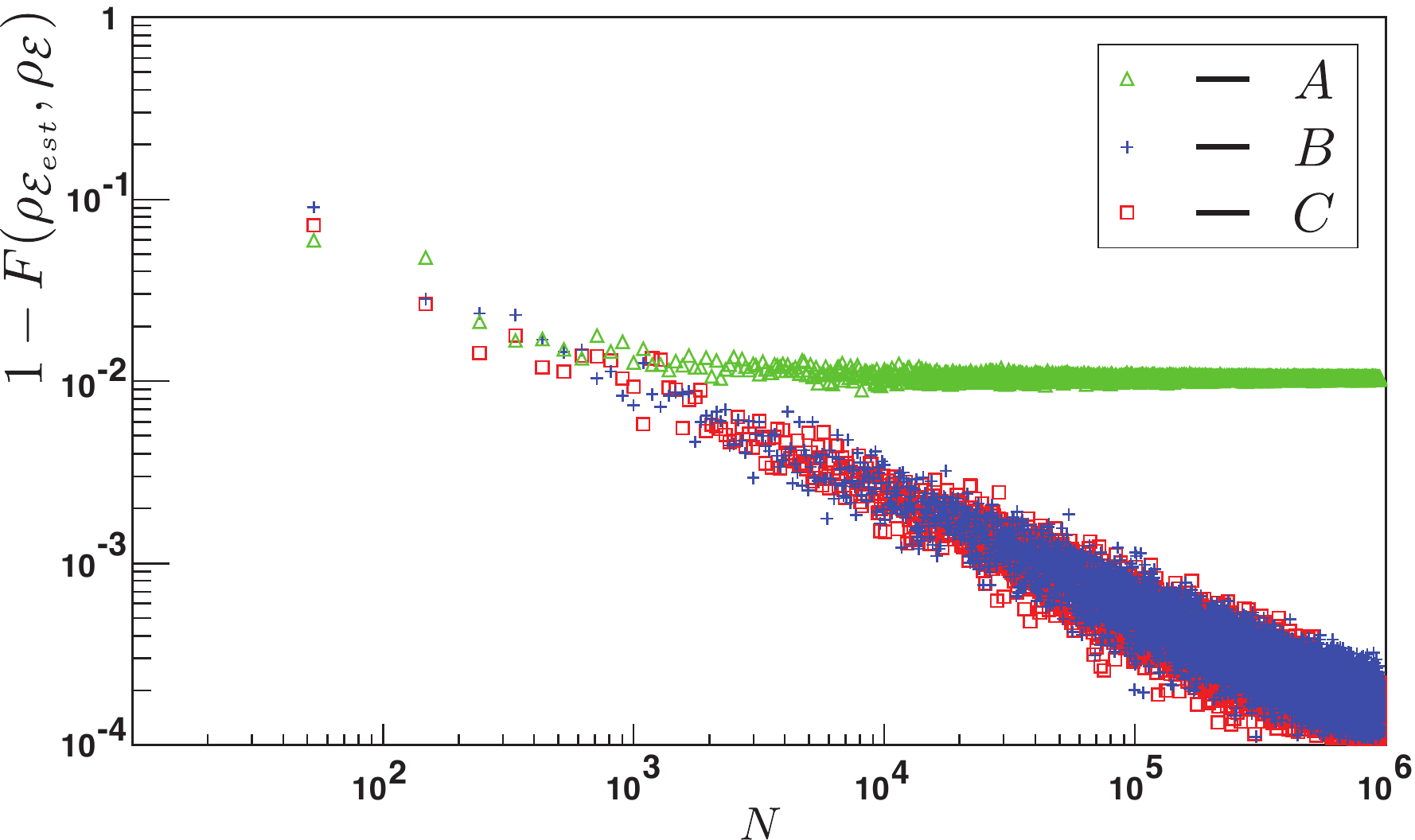}
	  \label{process_fidelity_true}
	}
	\caption{Process fidelities for the Hadamard gate, reconstructed using states and measurements characterised from method $A$ (green), $B$ (blue), and $C$ (red).  \subref{process_fidelity_ideal_sm1000},\subref{process_fidelity_ideal} the infidelity between the ideal unitary $\rho_{U}$ and the reconstructed process $\rho_{\m E_{est}}$ based on state and measurement tomography obtained for $N_{\rm SPAM} = 10^{6}$ \subref{process_fidelity_ideal_sm1000} and $N_{\rm SPAM} = 10^{9}$ \subref{process_fidelity_ideal} total measurements respectively.  The solid line represents the fidelity between the true process and the ideal Hadamard gate; process tomography estimates should converge to this line.  \subref{process_fidelity_true_sm1000},\subref{process_fidelity_true}  the fidelity between the true $\m E$ and reconstructed process $\m E_{\rm est}$, again for $N_{\rm SPAM} = 10^{6}$ \subref{process_fidelity_true_sm1000} and $N_{\rm SPAM} = 10^{9}$ \subref{process_fidelity_true} total measurements respectively.  These process tomography estimates should improve continuously as $1/\sqrt{N}$.  The data has been averaged over 10 runs per point.}\label{process_fidelity} 
\end{figure*}

We performed simulations of process tomography for the Hadamard gate, using the Choi matrix obtained by integrating a variant of the master equation~\eqref{masterqn} using a rotation axis corresponding to $J / \omega \approx 1$ for a period of time $t\approx \pi/\Omega$.  In a well-designed process tomography scheme, our estimate $\rho_{\m E_{\rm est}}$ should converge to the true state $\rho_{\m E}$ as $N \to \infty$.  However, we can expect that reconstruction errors in our estimates of the states and measurements used in this process tomography will affect this convergence to the true value.  This is borne out in our simulations; see Fig.~\ref{process_fidelity_true}.  Comparing the fidelity of the estimated state to the true state $F(\rho_{\m E_{\rm est}},\rho_{\m E})$ we see that all methods lead to process tomography that initially converges at a rate proportional to $N^{-1}$.  However, for large $N$ the errors do not converge to zero, but saturate at a nonzero lower bound determined by the accuracy of the state and measurement tomography.  For example, using $N_{\rm SPAM}=10^6$ to characterise the states and measurements (a value of $N_{\rm SPAM}$ for which the estimates obtained in method $A$ for state and measurement begin to saturate, see Fig.~\ref{comp_data}), we find that our process tomography stops improving after about $N \gtrsim 10^5$.  Note that this effect of the state and measurement errors on process tomography was also observed in Ref.~\cite{Merkel12}.  If, however, we perform state and measurement tomography with $N_{\rm SPAM} = 10^{9}$ (a value for which the estimates using method $A$ have long since saturated, but methods $B$ and $C$ are still improving), we see that the process tomography fidelities for methods $B$ and $C$ improve, converging as would be predicted, however method $A$ remains saturated at this limit.  This example clearly demonstrates the dependence of process tomography on state and measurement tomography. 

\section{Conclusion}

Quantum process tomography has become the gold standard for benchmarking quantum gates, in part because it constitutes a general method for characterising arbitrary processes without making unnecessary assumptions.  However, as applied in the past, substantial assumptions have been made about the form of the tomographically-complete set of states and measurements used for performing process tomography.  These assumptions, while possibly well-justified in optical and atomic systems, are inappropriate for most solid-state implementations.

We have presented and analysed numerically a range of methods for process tomography where these assumptions are relaxed or avoided altogether.  We showed that out of these methods, the one without any assumptions about the form of the tomographically-complete set of states and measurements (Method $C$) leads to the most efficient process tomography.  We note that our technique is related in spirit to the self-consistent tomography approach of Ref.~\cite{Merkel12}.  The key distinction between the two is that we first characterise our tomographically-complete states and measurements prior to initiating standard process tomography, with a relatively simple optimisation, whereas their approach performs a unified estimation of all gates, including those that could be used for state preparation and measurement in different bases, all at the same time.  Our work is relevant not just for the singlet-triplet qubit but for any system which has large SPAM errors. 

\begin{acknowledgments}
We thank Amir Yacoby, Hendrik Bluhm, Oliver Dial, Shannon Harvey, Michael Shulman, and David Reilly for discussions.  Research was supported by the Office of the Director of National Intelligence, Intelligence Advanced Research Projects Activity (IARPA), through the Army Research Office grant W911NF-12-1-0354.  We acknowledge support from the ARC via the Centre of Excellence in Engineered Quantum Systems (EQuS), project number CE110001013.
\end{acknowledgments}


\end{document}